\documentclass[]{emulateapj}
\usepackage{graphicx}
\shorttitle{Magnetic fields in CR shocks}
\begin{document}
\title{Turbulence-induced magnetic fields and structure of Cosmic Ray modified shocks}            
\author{A. Beresnyak\altaffilmark{1}, T. W. Jones\altaffilmark{2}, A. Lazarian\altaffilmark{1}}
\affil{Dept. of Astronomy, Univ. of Wisconsin, Madison, WI 53706}
\altaffiltext{1}{Dept. of Astronomy, Univ. of Wisconsin, Madison, WI 53706}
\altaffiltext{2}{Dept. of Astronomy, Univ. of Minnesota, Minneapolis, MN 55455}


\begin{abstract} 
  We propose a model for Diffusive Shock Acceleration (DSA) in which
  stochastic magnetic fields in the shock precursor are generated
  through purely fluid mechanisms of a so-called small-scale dynamo.
  This contrasts with previous DSA models that considered magnetic
  fields amplified through cosmic ray streaming instabilities; i.e.,
  either by way of individual particles resonant scattering in the magnetic
  fields, or by macroscopic electric currents associated with
  large-scale cosmic ray streaming.  Instead, in our picture, the
  solenoidal velocity perturbations that are required for the dynamo to
  work are produced through the interactions of the pressure gradient
  of the cosmic ray precursor and density perturbations in the
  inflowing fluid. Our estimates show that this mechanism provides fast
  growth of magnetic field and is very generic. We argue that for
  supernovae shocks the mechanism is capable of generating upstream
  magnetic fields that are sufficiently strong for accelerating cosmic
  rays up to around $10^{16}$ eV. No action of any other mechanism is necessary.
\end{abstract}


\section{Introduction}
Cosmic rays (CRs), relativistic charged particles with energies
$10^8-10^{22}$eV, constitute an essential part of astrophysical
systems (see Schlickeiser 2003). In galaxies they often provide
pressure and energy densities comparable to those of magnetic fields
and thermal gas. In very dense regions, such as the cores of molecular
clouds or accretion disks they are the only source of ionization that
that must be present to allow interaction of magnetic field and the fluid (see McKee \&
Ostriker 2007).  The origin of CRs has been a subject of debate from
the beginning of research in the field (Ginzburg \& Syrovatsky
1964).  By now, it is accepted, however, that 
galactic CRs at least up to the ``knee'' in the spectrum just
above $10^{15}$eV are most likely generated primarily by strong supernova shocks.

The modern study of CR acceleration in shocks involving so-called
diffusive shock acceleration (DSA) began with Krymsky (1977) and Bell
(1978) test particle models. They noted that when particles whose mean
free paths exceed the thickness of a viscous shock are scattered
upstream and downstream of the shock on ideal scatterers moving with
the bulk fluid, they gain energy each time they return across the shock
front. The resulting steady-state CR distribution is a power-law in
momentum, ($f \propto p^{-q}$) with a slope $q = 4 M^2/(M^2 - 1)$,
where $M$ is the shock flow Mach number.  This asymptotes to $q = 4$
for strong shocks.  It soon became apparent, if there is efficient
injection of fresh CRs at the shock, that CRs can extract significant
energy from the shocked flow. Since the CRs diffuse ahead of the
shock, this naturally leads to a pressure gradient upstream of the
shock transition that smoothly decelerates and compresses flow into
the shock, forming a shock precursor. In strongly modified shocks, the
total velocity jump through the precursor can exceed that
for a classical adiabatic fluid shock.
Since CRs with higher energies usually have longer mean
free paths, they travel further into the precursor than lower energy
CRs. Consequently, they ``see'' a stronger shock transition, and are
accelerated more efficiently.  This leads to an upward-concave spectrum
reaching a somewhat shallower slope than the test particle spectrum at
large energies (Malkov \& Drury 2001 and ref. therein).

On the other hand, the scattering events encountered by the CRs are
not ideal, but depend on some complex physics determined by details of
the local electromagnetic field, which is, in turn, modified by the
CRs. In a diffusive propagation approximation one must pay close
attention to this local physics to determine the spatial diffusion
coefficient, $D_{xx}$, and the momentum diffusion 
coefficient, $D_{pp}$, that go into the diffusion-convection equation for the
quasi-isotropic CR distribution function $f$:

\def\pder#1#2{\frac{\partial #1}{\partial #2}}

\begin{eqnarray}
\pder ft+u\pder fx & = &\pder{}{x}\left(D_{xx}\pder fx\right)\nonumber\\
                   & + & \frac p3 \pder ux \pder fp 
+ \frac 1{p^2} \pder{}{p}\left(p^2 D_{pp} \pder  fp\right),
\end{eqnarray}
(e.g., Skilling 1975), with some source term added for injection and
assuming that $f(x,p)$ depends only on one spatial coordinate $x$ and
the magnitude of CR momentum, $p$. This equation uses the so-called
``local'' system of reference, where the particle momentum is measured
with respect to the rest frame of the fluid. Similarly, magnetic field
perturbations that are the driver of particle scattering should also
be defined locally over the scales that are sampled by a particle
gyrating in the magnetic field\footnote{Fortunately, the modern theory of
  strong MHD turbulence, which uses a notion of ``critical balance''
  (Goldreich \& Sridhar 1995), is formulated in the terms of local,
  rather than global field (see Cho \& Vishniac 2000, Maron \&
  Goldreich 2001, Beresnyak \& Lazarian 2009a,b). How the turbulence may
  be modified in the presence of cosmic rays is still of an open
  question, however (see, e.g., Lazarian \& Beresnyak 2006).} (see,
e.g., Yan \& Lazarian 2004).
   
As the fluid and the high-energy particles couple through the
electromagnetic field, the problem reduces in practice to the study of
the generation of magnetic fields, their spatial structure and the
scattering of particles in those fields. This problem, formulated in
the most general case (e.g. with the fully three-dimensional Vlasov's
equation) is very hard to treat. In part this is due to the nonlinear
nature of the coupling between particles and fields. Following the
original test particle model a number of approaches to include more
complete physics have been adopted in the literature. One of the most
popular approaches to capture the feedback of CRs on the
electromagnetic fields has been to apply the streaming instability
mechanism, where particles, escaping upstream along the magnetic
field, confine themselves by amplification of resonant waves
(Lagage \& Cesarsky 1983). This approach assumes the existence of the
background magnetic field, which is usually taken for simplicity to be 
directed perpendicular to the shock; that is, along the shock
normal. This field was traditionally assumed to be of the order of the
background ISM magnetic field, $B_0\sim 5\mu G$.

However, it is easy to show that upstream magnetic fields of around
$5\mu G$ are too weak to provide an efficient acceleration of the
cosmic rays with energies as high as $10^{15}$ GeV, as required for
supernova shocks to produce CRs to the knee. PeV
cosmic rays have long mean free paths in such a field and have a
high probability of escaping the shock, so they are not subject to 
further acceleration. This poses a serious problem for the shock acceleration
of galactic CRs.

To overcome the problem one can argue that the magnetic field in the
preshock region can be much stronger than its interstellar value and
that the free energy available for the shock is sufficient to generate
much larger fields (Volk, Drury \& McKenzie, 1984). The magnetic field
generation, if pursued through streaming instability, leads naturally
to a highly nonlinear stage of the streaming instability where
$\delta B \gg B_0$. The original classical treatment of the
instability is not applicable in that limit. What happens in the non-linear regime
has been a subject of much discussion in recent years (e.g., Lucek \&
Bell 2000, Diamond \& Malkov 2007, Blasi \& Amato 2008, Zirakashvili
et al 2008, Riquelme \& Spitkovsky 2009). The current driven
instability proposed by Bell (2004) has moved recently to the center
of this scientific debate.  The driving electric current of that
instability comes from drift (streaming) of the escaping CRs. The {\it
  compensating} return current of the background plasma leads to a
transverse force on the background plasma that can amplify transverse
perturbations in the magnetic field. Numerical simulations suggest that
the initial field strength can grow substantially, leading in the
nonlinear form to disordered fields with coherence lengths that depend
on field strength (Bell 2004, Zirakashvili et al 2008, Riquelme \&
Spitkovsky 2009).

CR streaming instability in the presence of a relatively organized magnetic field
is an attractive starting point in understanding how CRs may amplify magnetic
fields. We argue, on the other hand, that the classic treatment of streaming
instability can not deal with $\delta B \gg B_0$ case, i.e., can not
generate large fields directly. This is because a large perturbed field
violates the key assumptions of particle dynamics made in
establishing the instability.  Various attempts have been made to
``renormalize'' the magnetic field so that the perturbation $\delta B$
becomes a new effective $B_0$. For instance, Diamond and Malkov (2007)
considered diffusion of magnetic energy from resonant gyroscales to
larger scales due to compressibility effects.  Among their assumptions
were weak coupling and advection-diffusion equations for wave number
densities\footnote{The assumption of weak coupling is broken in the
  case of strongly interacting waves, which necessarily appear in MHD
  turbulence (Goldreich \& Sridhar 1995), while the Fokker-Planck
  assumption is often broken even in the case of weak coupling, e.g.,
  when the interaction is mediated by wavevectors of comparable
  magnitude (Galtier et al 2000, Chandran 2005).}. Like most
treatments of this problem, their equations were also one-dimensional
in space, which required relatively simple structure of the cosmic ray
density with respect to magnetic field direction. The case of a highly
tangled three-dimensional magnetic field with $\delta B \gg B_0$, which
is the case we explore here, can not be treated in such way.

Indeed, in following a magnetic field line one will typically see
regions with alternating direction of the CR density gradient.
Therefore, streaming CRs will produce both forward and
backward going waves. This is unlike the classic picture where mostly
forward-going waves are produced. In a turbulent field those regions
of the alternating CR gradient will be non-trivially determined by the
topology of the field and the details of CR diffusion (see Appendix A).

Is it possible to generalize the streaming instability approach
to be applicable in a tangled magnetic field by
arguing that each particle scattering creates a magnetic field
perturbation on its own gyroscale? The difficulty with this is that in the tangled
$\delta B \gg B_0$ field such small incoherent perturbations are going to
average out, leaving the basic, residual effect that the CRs apply
pressure to the fluid. Can substantial magnetic fields be efficiently
generated in this generic case? This is the subject of the present
paper.

In what follows, we argue that there is an alternative process that
can provide fast magnetic field generation. It provides sufficiently
strong magnetic field without appealing to either classical streaming
instabilities or their modifications. The CR pressure gradient is the dynamical
agent that forms the shock precursor and also drives field amplification
\footnote{This physics clearly depends on prior development
of a strong CR precursor to the shock.
That requires waiting only until the shock is able to accelerate CRs
to relativistic energies (e.g., Kang et al. 2009). Even microGauss level fields,
without strong amplification, can do that quickly. For instance standard
diffusive shock acceleration using Bohm diffusion with a $1~\mu$G field in
a 10,000 km/s shock produces GeV CRs on a timescale of about a year.}.

We discuss below the generation of a strong magnetic
field as the precursor interacts with the interstellar
medium. We assume that the CR pressure is a smooth function applied to
the fluid, while the magnetic field is generated by
purely fluid nonlinear mechanisms. The magnetic field, in turn, plays
the role of CR scatterer and accelerator. 
The fluid is stirred within the precursor on large, precursor-sized,
scales by the combination of fluid density inhomogeneities and CR pressure.
Although we consider the CR pressure applied to the inflowing fluid
homogeneous in the direction along the shock, the acceleration of the
fluid element is highly inhomogeneous due to the density inhomogeneity
of the fluid. This drives precursor turbulence. 
The magnetic energy is generated on intermediate scales
through a small-scale dynamo. While the cosmic ray pressure gradient is
smooth, the generated magnetic field is tangled and effectively
suppress the streaming instabilities, at least in their classical
formulation. In assuming a smooth CR pressure distribution, we are
effectively averaging the CR action over small scales. We leave the
problem of the fluctuations of CR and magnetic pressure on gyroscale
to future study\footnote{In strongly CR modified shocks the
  acceleration of high energy particles mostly takes place within the
  precursor. Furthermore, in those shocks the CR pressure is
  predominantly in high-energy particles. As we will see in \S 3
  those particles do not have a gyroscale with the conventional
  meaning.}.
 
\begin{figure}
\plotone{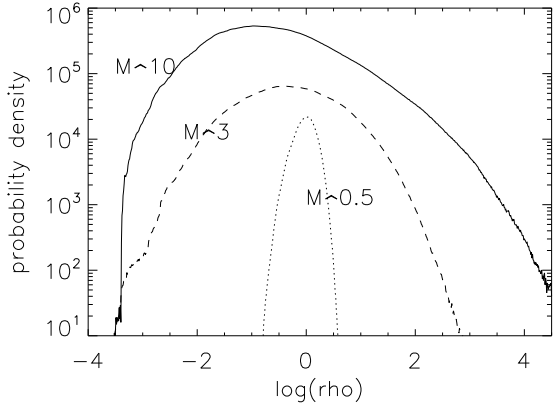}
\caption{PDF of density of the inflow fluid is approximately log-normal
due to the ambient turbulence in the ISM. Pictured is the PDF
of density from simulations with different sonic Mach numbers
(Beresnyak, Lazarian \& Cho, 2005).}
\label{pdf_rho}
\end{figure}

\section{Precursor-generated velocity field}
Although in a classic hydrodynamic shock no information can travel
upstream and create perturbations, the upstream diffusion of CRs leading
to precursor formation makes this possible, since CRs are much faster than the inflow.
Because CRs couple only diffusively to the fluid, their distribution
will generally be smoother and only slowly reactive to local changes in the fluid.
This creates a number of new possibilities, such as the Drury acoustic
instability (Drury 1984; Dorfi \& Drury 1985; Kang et al. 1992), which
is the enhancement of compressible perturbations by the CR pressure
gradient. Such instabilities, however, will only operate on
perturbations during the limited time that a fluid element crosses the
precursor.  This crossing time $\tau_c$ will be determined by the flow
profile as

\begin{equation}
\tau_c=\int^0_{x_0}\frac{dx}{u(x)}.
\end{equation}

On the other hand, generically, a fluid element passing through the
precursor has inhomogeneities of its own. Some level of density
inhomogeneity is always present in astrophysical plasmas, including 
the ISM (Armstrong et al. 1995) and stellar winds, media that
typically transmit SNR shocks.
These inhomogeneities, covering an extended
range of scales, are usually associated with MHD
turbulence\footnote{Turbulence arises due to the fact that microscopic
  dissipation coefficients such as viscosity and magnetic diffusivity
  are small and the Reynolds numbers are huge, which makes laminar
  flows in space a practical impossibility.}  (see, e.g., Elmegreen \&
  Scalo 2004, Lazarian \& Opher 2009).  Also there are so-called Small Ionized
and Neutral (SIN) structures, which exhibit fluctuations at the scale
from 100 to 1000 AU that are order of magnitude larger in 
amplitude than those
expected from the simplistic extension of the turbulent cascade to
small scales\footnote{The origins of these inhomogeneities are still
  debated.} (Heiles, 1997).  The stellar winds, acting as media
through which many young supernova shocks propagate, are also inhomogeneous and
turbulent.

Another generic property of the ISM and winds is cooling (radiative,
collisional, etc). Cooling keeps temperatures
fairly low, which makes ISM fluids pliable to compression.
The typical Mach numbers in the warm ISM are observed to be
between 1 and 10. Such strongly compressive flows are
characterized by significant nonlinear perturbations
of density\footnote{A
log-normal distribution of density is predicted and observed in simulations, see,
e.g., Beresnyak, Lazarian \& Cho (2005), for the supersonic isothermal magnetized
turbulence, see Fig.~\ref{pdf_rho}, while the subsonic case usually has a normal
distribution (which is a special case of log-normal)}. The hot ISM is
usually subsonic with respect to ambient turbulence, which suggests
smaller magnitude of the perturbations of density.

\begin{figure}
\plotone{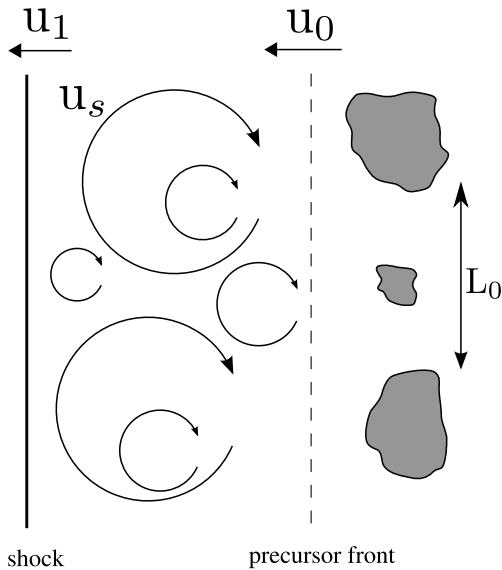}
\caption{Solenoidal motions, excited by CR precursor (the real picture is three-dimensional).
In the frame of the shock the preexisting perturbations enter the precursor creating
both compressive and solenoidal velocity perturbations (the last being depicted).}
\label{solen}
\end{figure}

As inhomogeneous fluid flows into the precursor with speed $u_0$ it is
gradually decelerated by the CR pressure gradient until it reaches the
speed $u_1$ at the dissipative shock front (Fig.~\ref{solen}). This
deceleration could create perturbations of velocity of the order $u_0-u_1$,
a difference between ballistic velocity of the high-density region and
full deceleration of the low density regions.
The resulting velocity field could not be purely divergent, but should be
partially solenoidal due to the coupling between compressible and
solenoidal motions on the outer scale of supersonic
turbulence\footnote{An easy way to visualize the effect of generation
  of the solenoidal motions is to consider the density inhomogeneities
  as random obstacles that interact with the flow. It is natural, that
  the flow get a solenoidal component as a result of such an
  interaction.}.  Simulations of strongly compressible turbulence
normally show a 2:1 ratio of solenoidal to compressive motions,
corresponding to 2:1 degrees of freedom in velocity. However,
this ratio was observed in simulations of statistically isotropic
stationary turbulence, while in our problem we have a preferred
direction, perpendicular to the shock front and a limited time for
the turbulence to develop. Taking into account our lack of knowledge
of the precursor highly compressible turbulence we parameterize
the amount of solenoidal motions with a parameter $A_s$ such
as the RMS of the solenoidal part of the velocity $u_s$ be defines as

\begin{equation}
u_s=A_s(u_0-u_1),
\end{equation}

where $A_s$ could be of the order of unity or smaller.
We designate the characteristic scale of solenoidal perturbations as $L$,
Normally, the supersonic ISM exhibits large density perturbations on a wide
range of scales. This is due to the fact that those are created mostly
by ISM turbulent slow sub-shocks. Although each secondary slow sub-shock
typically has a moderate Mach number ($\sim 2$), a random action of many
sub-shocks creates regions of high over-density or under-density
that can in rare instances exceed factors at least $10^4$ 
(see, e.g., Beresnyak, Lazarian \& Cho 2005).
More generally, the density perturbation from
individual sub-shocks is of order unity.  The characteristic scale here
is the distance between slow sub-shocks, which is smaller than the turbulence
outer scale, and could be of order of parsecs. From a practical point of
view, we will take $L$ from one of the typical scales of the problem,
either the above distance between sub-shocks or the thickness of the
precursor, or dynamical length $u_s\tau_c$, whichever is smaller
\footnote{Here we assumed that CR precursor is already mature,
i.e., the CR pressure is a considerable fraction of shock pressure
and the significant part of CR energy is in high-energy particles,
which makes precursor relatively thick. We leave the bootstrap
mechanism of the precursor to the future study.}.

In this paper we only consider turbulence hydrodynamically
amplified within the precursor.
We do not consider post-shock turbulence, which is a fairly well-known and
better explored phenomenon
that exists even in pure hydrodynamics without a CR precursor (see,
e.g., Giacalone \& Jokipii, 2007).  For the
purposes of this paper a rather simplistic description of the
post-shock magnetic fields is sufficient. It is well known that strong
fast shocks amplify magnetic field parallel to the shock front by,
approximately, the shock compression ratio $u_1/u_2$ (where $u_2$ is
the post-shock velocity).  Using that approximation we
will conservatively assume that the post-shock region has the same
magnetic field structure as the precursor, but with magnetic field
strength larger by a factor of $\sqrt{2/3} u_1/u_2$.

\section{The small-scale dynamo}
Three-dimensional solenoidal flows can amplify
magnetic fields through the stretch-fold mechanism.
This refers not only to turbulent flows, which have
motions on all scales, but even to large scale
three-dimensional laminar flows (e.g., Zeldovich et al, 1984).
In other words, amplification of magnetic fluctuations on small scales
is a very generic feature of highly-conducting fluids.  We will
consider a so-called generic small-scale dynamo (turbulent) in which
magnetic fields are amplified by initially weakly magnetized
hydrodynamic turbulence with an energy-containing (outer) scale of
$L$. ``Small-scale'' means that the scales of magnetic fields we are
interested in are smaller than $L$. The problem of the so-called
mean-field dynamo, when {\it large-scale} magnetic fields are
generated by small-scale motions is not considered here, because the
mean-field dynamo is fairly slow (see Vishniac \& Cho 2001), while inside
the shock precursor the time for the amplification is rather limited,
since all perturbations are quickly advected to the dissipative shock.
The small-scale dynamo has three principal stages -- a kinematic
stage, when magnetic energy grows exponentially, a linear stage
and a saturation stage (see Cho et al, 2009).

The kinematic stage of the dynamo has been studied extensively by
analytic (Kazantsev 1968, Kulsrud \& Anderson 1992) and numerical
tools.  The rising spectrum with magnetic energy $E_B(k)\sim k^{3/2}$
down to magnetic dissipation scales was predicted and later observed
in numerical simulations. For astrophysical applications the kinematic
dynamo is irrelevant, since its characteristic saturation timescale is
of the order of an eddy turnover time of the smallest eddies, which is
a tiny number compared to outer timescales. For our purposes we can
always assume that the kinematic dynamo is saturated and the dynamo is
in the linear stage.

\begin{figure}
\plotone{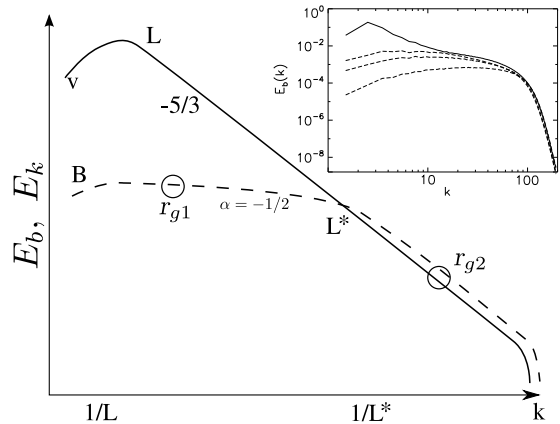}
\caption{Magnetic field spectrum (dashed), generated by a small-scale dynamo induced
by solenoidal velocity motions (solid). $L^*$ is an equipartition scale
of magnetic and kinetic motions, it plays a central role in particle
scattering. Upper panel: magnetic and velocity
fields from simulations (Cho et al 2009) with different dashed lines corresponding
to magnetic spectra at different times.}
\label{spectrum}
\end{figure}

In the linear stage magnetic energy grows linearly with time as

\begin{equation}
\frac{1}{8\pi}\frac{dB^2}{dt}=A_d \epsilon, 
\end{equation}

where $\epsilon$ is the energy transfer rate of the turbulence, which
can be estimated as $\epsilon=\rho u_s^3/L$, and $A_d$ can be called
an {\it efficiency of the small-scale dynamo}. A typical spectrum of
the velocity and the magnetic field in the linear stage is presented
on Fig.~\ref{spectrum}. At each particular time when the dynamo
operates, the magnetic field reaches equipartition with the turbulent
velocity field on some scale $L^*$.  This scale grows with time. On
scales smaller than $L^*$ magnetic and velocity perturbations form an
MHD turbulent cascade with a fairly steep spectrum. On scales larger
than $L^*$, the magnetic field has a fairly shallow spectrum and velocity
has a Kolmogorov spectrum.

The law of linear growth can be understood as follows. The main
cascade of energy is down-scale, but it is converted from a purely
velocity cascade to an MHD cascade at a scale $L^*$. One can imagine
that part of this energy cascades up (an inverse cascade) in the form
of magnetic energy. Let us call this fraction $A_d$. In principle,
$A_d$ can depend on scale, i.e., $A_d(L^*)$.  However, by an argument
similar to Kolmogorov's, if the inverse cascade mechanism is purely
nonlinear, then, in the middle of the inertial interval there is no
designated scale and, therefore, there is no dimensionless combination
involving $L^*$.  Therefore, the function $A_d(L^*)$ has to be
constant\footnote{This assumes locality of the small-scale dynamo. It
  is indirectly confirmed by the linear growth observed in
  simulations.}.  This gives a linear growth of energy.  The linear
growth can also be obtained if we assume that it takes {\it several}
turnover times to reach equipartition on each successive step to
larger and larger $L^*$ \footnote{Schekochihin \& Cowley 2007 proposed
  a model assuming that equipartition is reached at approximately one
  turnover time, which gives a linear growth of magnetic energy. This
  would correspond to our model with $A_d \sim 1$.}.  A linear growth
rate has been measured in Cho et al. (2009) as being close to
$A_d\approx 0.06$ which is the quantity we will use in this paper.

The shallow part of the magnetic spectrum between $L$ and $L^*$
normally has a slope $\alpha$ between $0$ and $-1$, as observed in
simulations. We will need these constraints later, when we describe a
model of particle scattering.  We particularly favor a model with
$\alpha=-1/2$. This model assumes that while magnetic fields on a
scale $L^*$ are generated by random eddies at the same scale $L^*$ and
contain most of the magnetic field energy, the larger scale fields
come from equipartition of magnetic tension on scale $l>L^*$. This can
be estimated as $\delta B^2(l)/l$, while the averaged magnetic tension
comes from a number $N=(l/L^*)^3$ of independent random eddies on
scale $L^*$.  This will give scalings $\delta B(l) \sim l^{-1/4}$ and

\begin{equation}
E_B(k)k=\delta B^2(k)\sim k^{1/2}.
\end{equation}

When $L^*$ approaches $L$, the small-scale dynamo enters the saturation stage
in which the magnetic field grows more slowly than in the previous linear stage.
The saturation value of magnetic energy depends slightly 
on the level of the mean magnetic field (Cho et al. 2009).
In our case the {\it mean} magnetic field can be considered negligible, as the
typical Alfv\'en velocity of warm ISM ($\sim 12$km/s) is
much smaller than the shock speed and associated turbulent
speeds (see \S~2). For the purpose of this paper, however, we won't need a
saturation stage, since we have limited time available for
amplification, $\tau_c$ (see \S~2),
which is normally not enough to reach the saturation stage\footnote{
Indeed, $L/u_s<\tau_c$ and saturation requires many $L/u_s$, since $A_d<<1$.}.  

From the linear stage growth we derive quantities $\delta B^*=\delta
B(L^*,x_1)$ and $L^*(x_1)$ that we will need in the next section:

\begin{equation}
\delta B^2(L^*,x_1)=8\pi A_d \epsilon \tau(x_1); 
\end{equation}

\begin{equation}
\frac{\delta B^*}{\sqrt{4\pi\rho}} = u_s\left(\frac{L^*(x_1)}{L}\right)^{1/3};
\end{equation}

and

\begin{equation}
\tau(x_1)=\int^{x_0}_{x_1}\frac{dx}{u(x)}; 
\end{equation}

\begin{equation}
L^*(x_1)=(2A_du_s\tau(x_1))^{3/2}L^{-1/2}.
\end{equation}

\section{Particle scattering and second-order acceleration}
In this section we derive $D_{xx}$ and $D_{pp}$ of the fast particles in the tangled
magnetic fields created by the small-scale dynamo. It should be noted that particle
dynamics is considered in the rest frame of the fluid (see also \S 1).

\begin{figure}
\plotone{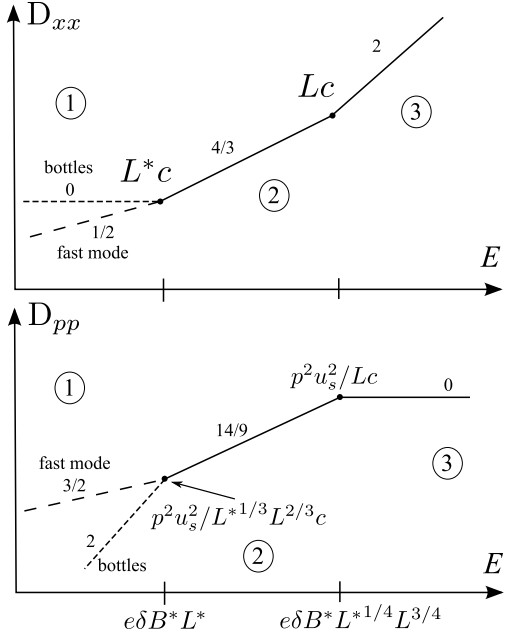}
\caption{Scattering coefficients as they depend on energy. The scattering has three
principal regimes: (1), low-energy scattering, which depends on the properties
of small-scale MHD turbulence (two cases where fast modes are present (dashed)
and absent (dotted) are shown); (2), strong scattering, where particles are
scattered efficiently by the strong magnetic fields generated by small-scale dynamo;
(3), high-energy scattering, where particles are only weakly scattered.}
\label{dcoeff}
\end{figure}

As it turns out, there are three different regimes of particle
scattering, depending on the particle energy, $E$, (see
Fig.~\ref{dcoeff}).  The magnetic spectrum, described in
Fig.~\ref{spectrum}, corresponds to the characteristic magnetic field
on a particular scale $\delta B(l)\sim\sqrt{E(k)k}$, which
increases\footnote{Provided that spectrum is sufficiently shallow,
  $\alpha>-1$} with decreasing scale as $l^{-\alpha/2-1/2}=l^{-1/4}$
for $\alpha = - 1/2$ from $L$ until $L^*$ and decreases with scale as $l^{1/3}$ for $l$
smaller than $L^*$.

If the particle energy is sufficiently low, the particle will, to the
first approximation, gyrate around a mean field.  Otherwise, its
trajectory will be stochastic.  In particular, if $E<<e\delta B(L^*)
L^*$, the particle will be gyrating along the mean field of $\delta
B^*=\delta B(L^*)$ with Larmor radius of $r_{g2}=E/e\delta B^*$.  We
will refer to these particles as low-energy and designate low energies
as region (1) in Fig.~\ref{dcoeff}.  For low energy particles the
scattering frequencies and acceleration will be determined by a
turbulence-based formulae with a turbulence outer scale of $L^*$ (the
scale with largest magnetic field).  We will consider this case in
detail in  \S 4.2.

\subsection{High energy particle scattering}
If the energy of the particle is higher than $e\delta
B(L^*) L^*$, there is no gyration and the particle's trajectory is
fairly stochastic. This is due to the fact that for this particle, the
vector magnetic field will be partially averaged out on larger scales
(so that $\delta B_l \sim l^{-1/4}$). Let us assume that such
a particle experiences a Bohm scattering and has a {\it mean free path} of

\begin{eqnarray}
r_{g1}&=&(E/e\delta B^*)^{\frac{2}{1-\alpha}}(L^*)^{-\frac{1+\alpha}{1-\alpha}}\nonumber\\
&=&(E/e\delta B^*)^{4/3}(L^*)^{-1/3}>L^*.
\end{eqnarray}

Indeed, (a) -- on such a scale, a particle will be deflected by an
angle of the order of unity; (b) -- the deflection from the larger scale
field will be smaller (if $l_2>r_{g1}$, then the deflection angle
$e\delta B(l_2) r_{g1}/E <1$); (c) -- the deflection from some smaller
scale $l_1$ is also smaller (it is a random walk with $r_{g1}/l_1$
steps), and the deflection angle is $e\delta B(l_1)
l^{1/2}r_{g1}^{1/2}/E <1$ \footnote{Provided that the spectrum is
  falling ($\alpha<0$)}).  Assuming $E=pc$ from here on, the $D_{xx}$ and $D_{pp}$
for such particles will be determined by velocity perturbations on
scale $r_{g1}$, i.e., $u_sr_{g1}^{1/3}L^{-1/3}$; thus,

\begin{eqnarray}
D_{xx}=(E/e\delta B^*)^{4/3}(L^*)^{-1/3}c,\\
D_{pp}=p^2u_s^2r_{g1}^{-1/3}L^{-2/3}/c\propto E^{14/9}
\end{eqnarray}

Finally, for very high energy particles, such as 

\begin{equation}
E>>e\delta B^* L^{\frac{1-\alpha}{2}}(L^*)^{\frac{1+\alpha}{2}}=e\delta B^* L^{3/4}(L^*)^{1/4},
\end{equation}

the trajectory will be a random walk with small deflections from scale 
$L$ and an effective mean free path of $E^2/e^2\delta B_L^2 L$. 
The $D_{pp}$ will be energy independent, as it will be set by
the outer scale velocity perturbations of $u_s$. Consequently,

\begin{eqnarray}
D_{xx}=E^2c/e^2\delta B_L^2 L,\\
D_{pp}=(e \delta B_L)^2 L u_s^2/c^3.
\end{eqnarray}

\subsection{Low energy particle scattering}
The scattering of gyrating particles in a strong mean field has been
studied for a long time (e.g., Jokipii, 1966), and an associated
so-called quasi-linear theory (QLT) of scattering has been formulated.
A particular property of MHD turbulence --- its strong anisotropy---
makes particle scattering from solenoidal modes (e.g., Alfv\'en) fairly
inefficient (Chandran 2000, Yan \& Lazarian 2002).  So, the
compressive fast mode, which is rather isotropic, has been proposed
as a more efficient scatterer in such settings (Yan \& Lazarian 2002,
2004). The fast mode, however, can be strongly damped in realistic
ISM environments, which causes scattering through the fast mode to depend
on the properties of the gas, such as temperature, density and
ionization fraction. For the purpose of this paper we ignore
complicated issues of low energy scattering and acceleration and
provide two simple, contrasting cases.

First, we can assume that the fast mode is fully damped and that
only solenoidal modes survive on scales of $L^*$ and smaller. 
In this case we can neglect QLT contributions from the Alfv\'enic and 
slow modes as they are very small. On the other hand, there are strong perturbations
of the magnetic field on the outer scale of $L^*$. In this case particles,
regardless of energy (provided that $E<<e\delta B(L^*)L^*$)
are going to be reflected by magnetic bottles
on scale $L^*$ and $D_{xx}$ will be independent of energy and equal to
$L^*c$, while $D_{pp}$ will be determined by the speed of the bottles. Thus,

\begin{eqnarray}
D_{xx}=L^*c,\\
D_{pp}=p^2u_s^2(L^*)^{-1/3}L^{-2/3}/c.
\end{eqnarray} 

Similar expressions can be derived from TTD resonance in the presence
of large magnetic field perturbations (Yan \& Lazarian 2008). Both of
these diffusion rates smoothly transition to the solutions from previous section.

Alternatively, we can assume that the fast mode is not damped and that
the scattering and second-order acceleration are due to the fast mode.
If we assume that the amplitude of the fast mode is approximately
in equipartition with other modes on the outer scale of sub-Alfv\'enic
turbulence, $L^*$, we will obtain expressions that also smoothly
transition to higher energy expressions from the previous section; namely,

\begin{eqnarray}
D_{xx}=cr_{g2}^{1/2}(L^*)^{1/2},\\
D_{pp}=p^2 u_s^2(L^*)^{1/6}L^{-2/3}r_{g2}^{-1/2},
\end{eqnarray}

(see, e.g. Yan \& Lazarian 2004). Here we have used so-called acoustic
turbulence scaling $\delta B\sim l^{1/4}$ for the isotropic fast mode,
as in Cho \& Lazarian (2002).

It is also possible that in the shock acceleration regions, where the
density of CRs is high, the scattering is affected by collective
effects where compressions of magnetic field induce the gyroresonance
instability in the fluid of compressed CRs as discussed in Lazarian \&
Beresnyak (2006). We do not provide a discussion of this more complex
case here.

\section{Heating}
Turbulent heating provided by solenoidal motions will be of the order
of the turbulent rate, $\epsilon=\rho u_s^3/L$. However, this assumes
that the turbulent cascade dissipates all of its energy into thermal
particles.  This might not be true for efficient second-order
acceleration of low-energy particles. In particular, if the spectrum
of CRs is sufficiently soft (steep), the second order acceleration will
drain energy from turbulence and put it into CRs as particles would
tend to diffuse to higher energies.  On the other hand,
if the spectrum is sufficiently hard (shallow), energy carried by high
energy CRs may be able to drive turbulence\footnote{This process
is qualitatively depicted in \S~2. The precursor can generate velocity fluctuations
due to inhomogeneity of the precursor pressure by a variety of mechanisms.}.  
In principle, one would like to monitor energy obtained or
lost by CRs due to second-order acceleration/deceleration and adjust
turbulent heating rates accordingly.

As a second note we mention that it is reasonable to believe that
dissipation from secondary shocks (similar to ISM turbulent slow shocks)
created in the precursor is going to be comparable to $\epsilon$.
The above considerations leave a certain degree of uncertainty in the
amount of thermal heating in our model.

\section{Discussion}

\subsection{Evolution of the ideas on shock acceleration}

The problem of nonlinear diffusive shock acceleration (DSA) is a
mature area of research with many publications since the original
linear theory papers of Krymsky (1977), Bell (1978), Axford et al
(1977) and Blandford \& Ostriker (1978). Most of this research was
motivated by the observed CR power-law distribution, and the apparent
robustness of the basic model.  The emphasis was largely, but not
exclusively, on the predicted properties of CRs, rather than the
physical details of their scattering.  So as long as the acceleration
process was described by rather simple and well-grounded
convection-diffusion equation (1), the details of scattering and fluid
dynamics have only had to be ``reasonable'' in order to obtain
applicable results.  On the other hand, since many reasonable models
of the detailed scattering physics were proposed, there were, in
effect, many models of DSA. One of the popular models to account for
the scattering is based on a linear ($\delta B< B$) or ``almost
linear'' ($\delta B \sim B$) streaming instability analysis.  The most
common application of the latter case is so-called Bohm scattering
(m.f.p. $\sim$ Larmor radius).  While these models result in
internally consistent particle spectra as well as the one-dimensional
structure of the flow (see, e.g., Malkov 1998; Berezhko \& Ellison
1999; Blasi 2002; Kang \& Jones 2007), the question of physical
justification remains.

Another approach has been to elicit some robust properties of the
acceleration process without regard to underlying particle-fluid
interactions (see a review of Malkov \& Drury 2001 and references therein),
although these studies emphasized the uncertainties in such important
quantities as the predicted high energy cutoff of accelerated
protons. This understanding is critical to a resolution of the origins
of galactic cosmic rays, and especially the so-called
``knee''. Furthermore, the higher energy frontier is more generally
important in astrophysics, as it could set limits to physics of what
is happening in such objects as AGNs or core-collapsed supernovae.
Indeed, while low-energy CRs can be accelerated in almost any source,
the highest energy CRs require a combination of large magnetic field
and large correlation length $B_ll$ to be contained in the source.
Otherwise, they easily escape. Present day neutrino experiments, such
as ICECUBE, have set an ambitious goal to peek into the hearts of these
objects, and obtain high-energy CR properties free of uncertainties
associated with models of escape and propagation.

Recently, the DSA problem has been reconsidered again with regard to
the problem of acceleration with more realistic scattering (Malkov \&
Diamond 2006) and the fluid dynamics in the presence of both the
strong streaming instability and strong compressibility (Diamond \&
Malkov 2007). The latter approach uses the advection-diffusion equation
with one spatial and one spectral dimension and assumes weak wave
coupling. This is unlikely to be enough to describe three-dimensional
compressive MHD turbulence (see the discussion in \S 1).  Also, we point
out that a classic streaming instability with a prescribed direction
of the CR gradient along magnetic field is unlikely to be useful in a
turbulent, strongly amplified field.  Although our model is at this point
phenomenological in its treatment of strongly compressible turbulence
and uses this compressibility to estimate solenoidal motions that
generate magnetic fields, we believe that we provided a more realistic
description of the magnetic fields generation in the preshocked gas.

Generation of the magnetic fields in the postshock region was
considered in many publications with both numerical and
phenomenological means (see, e.g., Cowsik \& Sarkar 1980,
Giacalone \& Jokipii 2007, Sironi \& Goodman 2007, Inoue et al, 2009).
Sironi \& Goodman (2007) estimated a vortical energy that
appears after GRB afterglow shocks due to preexisting density
inhomogeneities. However it assumed that the magnetic field reaches
equipartition with vortical energy and did not discuss the structure
of the magnetic field.  We argue that it is the shock precursor fields
which are important. Also, the spectrum and the structure of this field
are paramount for understanding physically motivated scattering
coefficients. Inoue et al (2009) provided a detailed two-dimensional
numerical study of post-shock turbulence and dynamo action. Two-dimensional dynamics,
however is very different from a three-dimensional one (see, e.g., Biskamp 2003).

\subsection{Current limitations of the numerical approach}
  
The full numerical treatment of the highly-compressible flows in question is
difficult. One-dimensional studies of such flows are meaningless, as
they are likely to produce a picture which is completely different
from three-dimensional dynamics. Indeed, one-dimensional flows
necessarily generate shocks in a finite time, and the dynamics are
fully dominated by those shocks (see, e.g. Suzuki et al 2007). The
three-dimensional dynamics is more complicated, with compressible
motions containing only a fraction of energy and weak sub-shocks playing some,
but not necessarily a dominate role. Direct three-dimensional simulations
of supersonic turbulence are not only computationally expensive, but
inherently limited in a number of ways.  The  best known
limitation is the range of scales (typically a useful range of
scales in a fairly computationally expensive $1024^3$ simulation is
4-200 in grid units). More relevant to the DSA problem, however, is
the limitation in sonic Mach number, which is around 20 for a $1024^3$
simulation and is determined by having strongly compressible motions
($v\sim c_s$) on the grid cell scale, or having a significant fraction
of matter accumulated in clumps of the grid cell size. The requirement
of the DSA problem, however,
is not $M_s\sim 20$ but rather can be $M_s\sim 1000$ or even higher
in more energetic sources (AGNs, relativistic jets in GRB sources).
In this situation three-dimensional fluid-PIC codes that will aim
to describe interacting fluid and CRs will suffer from the same
limitations as fluid codes, but also will be unable to describe
a huge spread in energies of the CR spectrum, due to limited statistics
of particles.

\subsection{The source of solenoidal motions}
In \S 2 we assumed that the solenoidal velocity is a fraction
of the velocity drop along the precursor with the main mechanism
being the pre-existing density inhomogeneities and the precursor
pressure field. We noted that we expect this mechanism to be fairly
efficient (i.e. $A_s$ close to unity) as long as preexisting
density perturbations $\delta \rho/\rho$ are of the order unity.
However, it is interesting to consider the possibility that
the initial density perturbations are enhanced {\it by
the precursor} to this level. For instance, once the total pressure in the
precursor, $P$, is dominated by CRs, the weak coupling between
fluid fluctuations and CRs leads to regions in the
precursor where $\nabla P \cdot \nabla \rho < 0$. This is
unstable to Rayleigh-Taylor-like instabilities that have been shown to
strongly enhance turbulence in CR modified shocks (Ryu et al. 1993).
While these effects have been demonstrated, the resulting solenoidal
turbulence has not yet been quantitatively evaluated.
Therefore, we cannot conclude that $A_s$ could be considered a small
parameter even if the inflowing density perturbations are small.

Aside from inflowing density inhomogeneities, another possible mechanism of
generating turbulence is related to the inhomogeneities in the precursor
CR density, or, more generally, a dynamic three-dimensional turbulent
interaction between fluid and CR's, which allows exchange of energy in
both directions. This includes instabilities discussed above.
Qualitatively, this effect also works to create additional solenoidal
motions. However, this is a more complicated
phenomenon that will be studied elsewhere.

We expect the current instability, which is considered in detail
in the next subsection, to generate density and solenoidal
velocity perturbations as well. However, due to the limitations
of the numerical approach (see the previous subsection)
the dynamics of density in the nonlinear stage of current instability
is not fully understood.

\subsection{Our approach and current driven instability}

In response to the realization that magnetic fields could
be amplified significantly compared to the background field,
a model based on a current-driven instability was
proposed and tested numerically (Bell 2004, Vladimirov et al. 2006;
Zirakashvili et al 2008, Riquelme \& Spitkovsky 2009). In the linear
instability stage the magnetic field grows fastest on the
characteristic scale, determined by the initial field $B_0$, and the
current $j_d$,

\begin{equation}
l=1/k_c=\frac{cB_0}{4\pi j_d},
\end{equation}

where $j_d$ is from high-energy CRs that are ``rigid'' enough
to have $qB_0/pc<<k_c$ (Bell 2004). 
The linear growth rate depends only on the current according
to the relation
\begin{equation}
\gamma =\frac{j_d}{c\sqrt{\frac{\rho_0}{\pi}}}.
\end{equation}

The nonlinear saturation stage is characterized by slower growth on
larger scales (Bell 2004, Zirakashvili et al 2008).  This growth,
however, advects along with the fluid and, as we discussed in
\S~2, has to be limited by the time for flow to cross the precursor.
It is also worth noting that Bell (2004) and Zirakashvili et al (2008)
used weakly compressible simulations with initial conditions that did
not have strong perturbations in either fluid density or CR density,
which makes it totally different from our approach. We believe, there
is a good physical reason why precursor turbulence is often
strongly compressible and inhomogeneous (see \S 2).

One can compare growth rates of magnetic energy, provided by the
small-scale dynamo (eq. 8) and current-driven instability (eq. 21),
assuming that the current-driven instability is in the stage with
$\delta B\sim B_0$, but the linear growth rate still holds.
The result is

\begin{equation}
\frac{dB^2_{cur}}{dB^2_{dyn}}=52\cdot\frac{j_dL}{cB_0}\cdot\left(\frac{v_{A0}}{u_s}\right)^3.
\label{9}
\end{equation}

The second part of the RHS in eq. (\ref{9}) can be estimated from the characteristic 
$v_A$ of the ISM ($\sim 10$km/s) and could be as small as $10^{-9}$, if the shock speed
is large ($\sim 10000$km/s).
The first part of the RHS of eq. (\ref{9}) can be interpreted as the ratio of the 
field created
directly by the high-energy particle current on scale $L$ to the initial field $B_0$. 
The robust estimate of the current $j_d$, however, seems pretty elusive.
Suppose, following Riquelme \& Spitkovsky (2009),
we assume that the current is produced entirely by escaping
particles and that there is a fixed ratio $\eta_{esc}\approx 0.05$ between
the flux of CR energy emitted by the shock and the flux of
energy of the incoming fluid $\rho u_{sh}^3/2$ and also assume
that a characteristic energy of escaping particles is $E_{esc}=10^{15}$ eV.
Then one can numerically estimate the above ratio.
Taking $L=1$pc and assuming $u_s\approx 0.5 u_{sh}$ (since we assume
that the shock is strongly modified, i.e. $u_{sh}=u_0\sim u_0-u_1$ and $A_s$ is
of the order unity), we get

\def\bratio#1#2{\left(\frac{#1}{#2}\right)}

\begin{eqnarray}
\frac{dB^2_{cur}}{dB^2_{dyn}}&=&1.6\times 10^{-4}\bratio{10^{15}\mbox{eV}}{E_{esc}}
\bratio{\eta_{esc}}{0.05}\bratio{L}{1\mbox{pc}}\nonumber\\
&\times&\bratio{B_0}{5\mu\mbox{G}} \bratio{v_{A0}}{12\mbox{km/s}}\bratio{0.5 u_{sh}}{A_s(u_0-u_1)}^3.
\end{eqnarray}

This difference in field growths is due to the fact that in our model
the full pressure ($\sim$ energy density) of the CRs is behind the
force that winds up magnetic fields, while the driving force of the
current instability model comes only from those CRs that are able
to freely stream a substantial distance through the flow.
The efficiency of our model is based on an assumed large value
of $A_s$ parameter (of the order unity). Also we do not consider
the feedback of CRs to the full three-dimensional fluid dynamics,
assuming instead that CR pressure is homogeneous in two directions
along the shock.

\subsection{Future work}

The self-consistent treatment of the flow profile and acceleration of
particles using expressions from \S~4 will be presented in a future
publication. The scattering coefficients assume that efficient
scattering and acceleration will be provided for particles with
energies up to $E_{2-3}=e\delta B^* {L^*}^{1/4}L^{3/4}$ (see Fig.~\ref{dcoeff}).
This energy can be estimated taking $u_s\tau\approx L$, $u_s\approx
10^4 \mbox{km/s}$ and $L\approx 1$pc as $E_{2-3}\approx
3\times10^{17}$ eV. This energy corresponds to a mean free path of the
order of $L$. However, as the acceleration efficiency is smaller by a factor
of $u_s/c$ (Hillas, 1984) the maximum acceleration energy will be
around $10^{16}$ eV. The higher energy particles will be scattered
relatively less efficiently and likely to form a steeper spectrum.
This estimate is tentative and needs to be confirmed or corrected
as the self-consistent treatment of Eq. (1) and the evolving
structure of the precursor and will be available from the future work.

Finally, we would like to mention that field growth in our model could
be amended by the consistent description of the front-running region of
the precursor where turbulence is not yet developed.  At this point,
however, it is not clear whether this region will be dominated by
the classic streaming instability, strong compressible effects (and,
possibly, generation of secondary fast shocks, as the magnetic field
is not yet amplified to prevent creation of those), or the
current-driven instability.

\section{Summary}
In order to explain efficient acceleration of high-energy CRs in
supernova shocks we appealed to magnetic field amplification in the
shock precursor that is induced by the small scale turbulent dynamo. In
our picture the velocity field necessary for such amplification
appears hydrodynamically as a result of the strong CR pressure
gradient acting on an initially {\it inhomogeneous} medium in the
preshock region. We assumed efficient conversion of the precursor
velocity drop into solenoidal motions due to strong density
perturbations of the inflowing fluid. We also ignored CR inhomogeneities
appearing as a back-reaction and their nonlinear feedback.
We estimate that magnetic fields produced by such
amplification are able to efficiently scatter and accelerate
CRs with energies up to $10^{16}$ eV.

\acknowledgments
We are grateful to Misha Malkov, Pat Diamond, Anatoly Spitkovsky
and Brian Reville for fruitful discussions. We are grateful to the anonymous
referee for helpful, meticulous comments.
AB thanks the IceCube project for support of his research.
TWJ acknowledges support from NASA grants NNG05GF57G and NNX09AH78G, as
well as the Minnesota Supercomputing Institute. AL acknowledges the NSF grants
ATM 0648699, AST 0808118 and the NASA grant NNX09AH78G. 

\section{Appendix A}
In order to demonstrate that the stochastic field generated in
the cosmic ray precursor will be hard to treat properly
in the classic streaming
instability approach, we ran particle tracing simulations in
a stochastic three-dimensional field, generated by a small scale
dynamo (discussed in \S 3).

\begin{figure}
\includegraphics[width=\columnwidth]{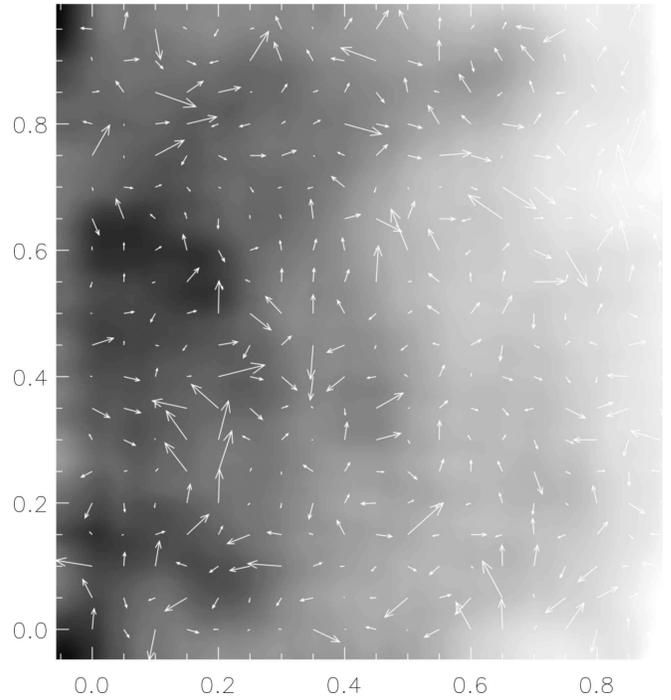}
\caption{A 2D slice of the 3D particle tracing experiment, showing magnetic
field projection on the plane (arrows) and particle density using
grayscale (dark being higher density).}
\label{cr_demo}
\end{figure}

The electromagnetic field was obtained through direct three-dimensional
numerical simulation
of the incompressible MHD equations with turbulent driving, along with
a zero mean field and small fluctuations as initial conditions. The simulation
was similar to MHD2b0h, described in Beresnyak \& Lazarian (2009b), with the exception
that it was run for a relatively short time while the small-scale 
dynamo was in its
linear stage (see \S 3). The time was chosen so that the equipartition scale $L^*$
was approximately in the middle of the logarithmic range of scales. The electric
field was obtained assuming $v_A/c=10^{-5}$.

The particles were injected on one side of the cube and their relativistic equation
of motion was solved by a hybrid quality-controlling Runge-Kutta ODE solver.  
The particles were injected from the left and escaped to the right.
Fig.~\ref{cr_demo} presents a slice of this three-dimensional experiment,
where magnetic fields are represented by arrows and the CR particle density by grayscale.
We see, that aside from the obvious left-to-right global gradient we cannot identify any
clear structure of particle density along the field. This invalidates
in this case the assumption of the classic streaming instability, which
requires a regular particle density gradient along the field lines to produce
a regular field-aligned number density current.

\end{document}